\documentclass{aastex}
\usepackage{emulateapj5,apjfonts,epsfig}
\usepackage{amstext}
\usepackage{amsmath}

\slugcomment{Accepted for publication in the Astrophysical Journal}
\shorttitle{SPECTROSCOPY OF XTE J0929--314}
\shortauthors{JUETT, GALLOWAY, \& CHAKRABARTY}

\begin{document}

\newcommand{\J}{XTE~J0929$-$314}
\newcommand{\XTE}{{\em RXTE}}
\newcommand{\Ch}{{\em Chandra}}
\newcommand{\eps}{{\rm erg\,s^{-1}}}
\newcommand{\epcs}{{\rm erg\,cm^{-2}\,s^{-1}}}
\newcommand{\cts}{{\rm count\,s^{-1}}}

\title{X-ray Spectroscopy of the Accreting Millisecond Pulsar
XTE~J0929--314 in Outburst}

\author{Adrienne~M.~Juett\altaffilmark{1}, Duncan K. Galloway, and
Deepto~Chakrabarty\altaffilmark{1,2}}

\affil{\footnotesize Center for Space Research, Massachusetts
Institute of Technology, Cambridge, MA 02139;\\ ajuett, duncan,
deepto@space.mit.edu}

\altaffiltext{1}{Also Department of Physics, Massachusetts Institute
of Technology, Cambridge, MA 02139} 
\altaffiltext{2}{Alfred P. Sloan Research Fellow}

\begin{abstract}
We present the high-resolution spectrum of the accretion-powered
millisecond pulsar \J\/ during its 2002 outburst, measured using the
Low Energy Transmission Grating Spectrometer onboard the {\em Chandra
X-ray Observatory}.  The 1.5--25.3~\AA\/ (0.5--8.3~keV) \Ch\/ spectrum
is well fit by a power-law $+$ blackbody model with photon index
$\Gamma=1.55\pm$0.03, blackbody temperature $kT_{\rm
bb}=0.65\pm$0.03~keV, and blackbody normalization $R_{\rm
bb,km}/d_{\rm 10kpc}=7.6\pm0.8$.  No emission or absorption features
are found in the high-resolution spectrum, with a 3$\sigma$ equivalent
width upper limit of $<0.007$~\AA\/ at 1.5~\AA\/ and $<0.12$~\AA\/ at
24~\AA.  The neutral absorption edge depths are consistent with the
estimated interstellar absorption along the line of sight to the
source.  We found no orbital modulation of the 2--10~keV X-ray flux,
to a 3$\sigma$ upper limit of 1.1\%, which implies an upper limit on
the binary inclination angle of $i\lesssim 85^{\circ}$ for a
Roche-lobe--filling companion.  We also present the broadband spectrum
measured over the course of the outburst by the {\em Rossi X-ray
Timing Explorer} (\XTE).  The \XTE\/ spectrum of \J\/ is also well fit
with a power-law $+$ blackbody model, with average values of
$\Gamma=1.76\pm$0.03, $kT_{\rm bb}=0.66\pm$0.06~keV, and $R_{\rm
bb,km}/d_{\rm 10kpc}=5.9\pm1.3$ in the 2--50~keV energy range.  The
blackbody flux remained constant over the course of the outburst,
while the power-law flux was strongly correlated to the (decreasing)
flux of the source.  We find that the difference in power-law photon
indices measured from \Ch\/ and \XTE\/ spectra can be explained by a
change in the power-law photon index at low energies.
\end{abstract}

\vspace{0.1in}

\section{Introduction}
Millisecond pulsars (MSPs) have long been considered one of the
possible endpoints of low-mass X-ray binary (LMXB) evolution.  The
neutron star (NS) is thought to be spun-up to a millisecond period by
accretion from its low-mass companion.  After the accretion phase has
ended, the NS may turn on as a radio MSP.  In the last five years,
this theory has been confirmed with the identification of three
accretion-powered MSPs, SAX J1808.4$-$3658, XTE~J1751$-$305 and \J\/
\citep{wv98,cm98,mss+02,gcm+02}.

Interestingly, all three accretion-powered MSPs are in short period
binaries, with the two most recently discovered having binary periods
$\approx$43 min \citep{mss+02,gcm+02}.  These short periods place the
MSPs XTE~J1751$-$305 and \J\/ in the class of ultracompact binaries,
defined as having orbital periods $\lesssim$80 min.  Ultracompact
binaries require hydrogen-deficient or degenerate donors
\citep*{nrj86}.  Recently, O and Ne emission and absorption features
were discovered in two known and three suspected ultracompact systems
\citep*[][]{scm+01,jpc01,jc02}.  These results led the authors to
conclude that the donor stars are degenerate C-O WDs.  On the other
hand, the {\em XMM-Newton} spectrum of XTE~J1751$-$305 did not show
emission or absorption features \citep{mwm+02}.  In addition, no
unusual abundances were required to fit the neutral absorption edges.
This result is consistent with the suggestion of \citet{mss+02} that
the donor in XTE~J1751$-$305 is a He WD \citep[see also,][]{b02}.

\J\/ was discovered in April 2002 by the All Sky Monitor onboard the
{\em Rossi X-ray Timing Explorer} \citep[\XTE;][]{r02}.  Further
\XTE\/ observations detected 185~Hz pulsations modulated by a 43.6-min
binary orbit \citep*{rss02,gcm+02}.  Radio and optical counterparts
with positions consistent with the X-ray position were also detected
\citep*{rdm02,ggh02,c02}.  An optical spectrum of \J\/ revealed
emission lines from \ion{C}{3}/\ion{N}{3} $\lambda$4640-4650 and
H$\alpha$ $\lambda$6563 \citep{ccg+02}.  Given the optical detection
of emission lines, and the ultracompact nature and high-Galactic
latitude of the source, \J\/ is an ideal candidate for a
high-resolution X-ray spectroscopic study to search for emission and
absorption features similar to those seen in other ultracompact
systems.  In this letter, we present results from a Director's
Discretionary Time observation of \J\/ with the {\em Chandra X-ray
Observatory}, as well as spectral results from the \XTE\/ pointed
observations throughout the outburst.

\vspace{0.2in}
\section{Observation and Data Reduction}
The 2002 outburst of \J\/ began around MJD~52370 and peaked at
$\approx~31$~mCrab (2--10~keV) on MJD~52394 \citep{gcm+02}. The
pointed \XTE\/ observations commenced 2~d later, and proceeded once
every few days through MJD~52456.  The source dropped below the
$3\sigma$ detection limit of $7.5\times10^{-12}\ \epcs$ (2--10~keV;
equivalent to 0.32~mCrab) after MJD~52443.

\J\/ was observed throughout its 2002 outburst by the two pointed
instruments onboard \XTE.  The Proportional Counter Array
\citep[PCA;][]{xte96} consists of five gas-filled proportional counter
units (PCUs) with a total effective area of $\approx6000\ {\rm cm}^2$,
sensitive to X-ray photons in the 2.5--60~keV range.  The two
instrument clusters comprising the High-Energy X-ray Timing Experiment
\cite[HEXTE;][]{hexte96} present an effective area of $\approx1600\
{\rm cm}^2$ to photons in the energy range 15--250~keV. We extracted
average spectra from standard instrument mode data (``Standard-2'' and
``Archive'' for PCA and HEXTE, respectively) from each PCU/cluster for
each observation. Since the instrumental gain is known to vary between
PCUs, and also with time, we generated a separate response matrix for
each PCU and each observation using {\tt pcarsp} version 8.0, supplied
with LHEASOFT version 5.2. We estimated background spectra using the
``combined'' gain epoch 5 (beginning 2000 May 13) faint-source model
as input to {\tt pcabackest} version 3.0.  Over the course of the
outburst, we collected 38 observations of \J\/ totaling 123.5~ks.

We also observed \J\/ with \Ch\/ on 2002 May 15 (MJD 52409) for 18 ks
using the Low Energy Transmission Grating Spectrometer (LETGS) and the
Advanced CCD Imaging Spectrometer \citep[ACIS;][]{wbc+02}.  The LETGS
spectra are imaged by ACIS, an array of six CCD detectors.  The
LETGS/ACIS combination provides both an undispersed (zeroth order)
image and dispersed spectra from the grating with a first order
wavelength range of 1.4--63~\AA\/ (0.2--8.9~keV) and a spectral
resolution of $\Delta\lambda=$ 0.05~\AA.  The various orders overlap
and are sorted using the intrinsic energy resolution of the ACIS CCDs.
The observation used a Y-offset of $+$1\farcm5 in order to place the
O-K absorption edge on the back-side illuminated S3 CCD, which has
suffered less degradation than the front-side illuminated CCDs.

Using {\tt tgdetect}, we determined the zeroth order source position
of \J\/ to be: R.A.=$09^{\rm h} 29^{\rm m} 20\fs15$ and
Dec=$-31^{\circ} 23\arcmin 04\farcs3$, equinox J2000.0 (90\%
confidence error of 0\farcs6\footnote{See
http://asc.harvard.edu/cal/ASPECT/celmon/index.html}).  The \Ch\/
position is consistent with both the optical and radio counterpart
positions \citep{rdm02,ggh02,c02}.  The first order dispersed spectrum
of \J\/ had an average count rate of 7.64$\pm$0.02~counts~s$^{-1}$.
We examined the total count rate, as well as the count rates in two
different energy ranges, to check for changes in the spectral state.
We found no evidence for any change of spectral state during the \Ch\/
observation.

The ``level 1'' event file was processed using the CIAO v2.2 data
analysis package\footnote{http://asc.harvard.edu/ciao/}.  The standard
CIAO spectral reduction procedure was performed.  We filtered the
event file retaining those events tagged as afterglow events by the
{\tt acis\_detect\_afterglow} tool.  Since order-sorting of grating
spectra provides efficient rejection of background events, the
afterglow detection tool is not necessary to detect cosmic ray
afterglow events.  No features were found that might be attributable
to afterglow events.

For bright sources, pileup can be a problem for CCD detectors
\citep[see, e.g.,][]{d02}.  The zeroth order \Ch\/ spectra of \J\/ was
heavily affected by pileup and was not used in this analysis.  In
addition, it was found that the first order spectrum of \J\/ suffered
from pileup in the range 5--12~\AA\/ (1--2.5~keV).  In order to use
the grating pileup kernel in ISIS, we created detector response files
(ARFs) for each chip using the CIAO tool {\tt mkgarf} and then
combined them to create $+1$ and $-1$ order ARFs using a custom tool
developed by J. Davis of the HETGS instrument team.  (This tool is
similar to the standard CIAO tool but correctly calculates the
fractional exposure at each response bin.  This functionality has been
added to the standard tools in CIAO v2.3.)  The grating pileup kernel
models the effect of pileup on grating spectra, similar to the pileup
model available for CCD spectra in ISIS and XSPEC \citep[see,][for a
discussion of grating pileup modeling]{d02}.  In addition, the data
and responses were rebinned to 0.083~\AA\/ to reflect the size of an
ACIS event detection cell (3 CCD pixels).  Background spectra were
extracted from the standard LETG background regions.  Spectral
analysis of the \Ch\/ observation of \J\/ was performed using ISIS
\citep{hd00}.

\vspace*{0.1in}
\section{Search for Orbital Modulation}
The presence (or absence) of orbital modulation of the X-ray flux can
help constrain the inclination of the binary, and in turn the mass of
the companion.  \citet{gcm+02} determined the orbital period of \J\/
to be 43.6~min from the Doppler modulation of the pulse arrival times.
We searched both the \XTE\/ and \Ch\/ data for orbital modulation of
the X-ray flux.

For the \XTE\/ analysis, we calculated background-subtracted,
solar-system barycenter corrected, 16-s binned lightcurves for each
observation from photons in the energy range 3--13~keV.  This range
was selected to maximize the signal-to-noise ratio, and is identical
to that used for the pulse timing analysis of \citet{gcm+02}.  For
each observation, we created a phase folded lightcurve of 8 bins using
the measured orbital parameters, with phase zero set to the epoch of
$90^{\circ}$ mean longitude, $T_{\pi/2}$.  The length of each
observation was short compared to the decay timescale of the X-ray
flux of \J, so it was not necessary to correct for the decline in
X-ray flux during a single observation.  Each lightcurve was then fit
with a sine curve with variable amplitude and phase.  From these
results, we place a 3$\sigma$ upper limit on the fractional rms
modulation of 1.1\%.

We also searched the \Ch\/ data for orbital modulation of the X-ray
flux.  For the \Ch\/ analysis, we first barycentered and randomized
the first order grating events.  Randomizing of the event arrival
times consists of adding a random quantity uniformly distributed
between 0--3.2 s, in order to avoid aliasing caused by the readout
time.  We calculated the phase of each first order event using the
orbital parameters from \citet{gcm+02} and created an 8 bin
phase-folded lightcurve.  We found an upper limit of 2.3\% rms,
consistent with the \XTE\/ results.

Since the \Ch\/ data is not subject to time gaps due to spacecraft
motion, we also calculated a Fourier transform of the first order
events which had been made into a 10-s binned lightcurve.  We searched
for modulations of the X-ray flux with frequencies between
6$\times10^{-5}$ and 5$\times10^{-2}$ Hz.  We found no evidence for
any periodic modulation, with a 90\%-confidence upper limit of 1.4\%
for the fractional rms amplitude over the frequency range.  The
90\%-confidence upper limit for the fractional rms amplitude of a
signal at the orbital period is 0.4\%, well within the upper limits
given by the phase folding analyses.

\centerline{\epsfig{file=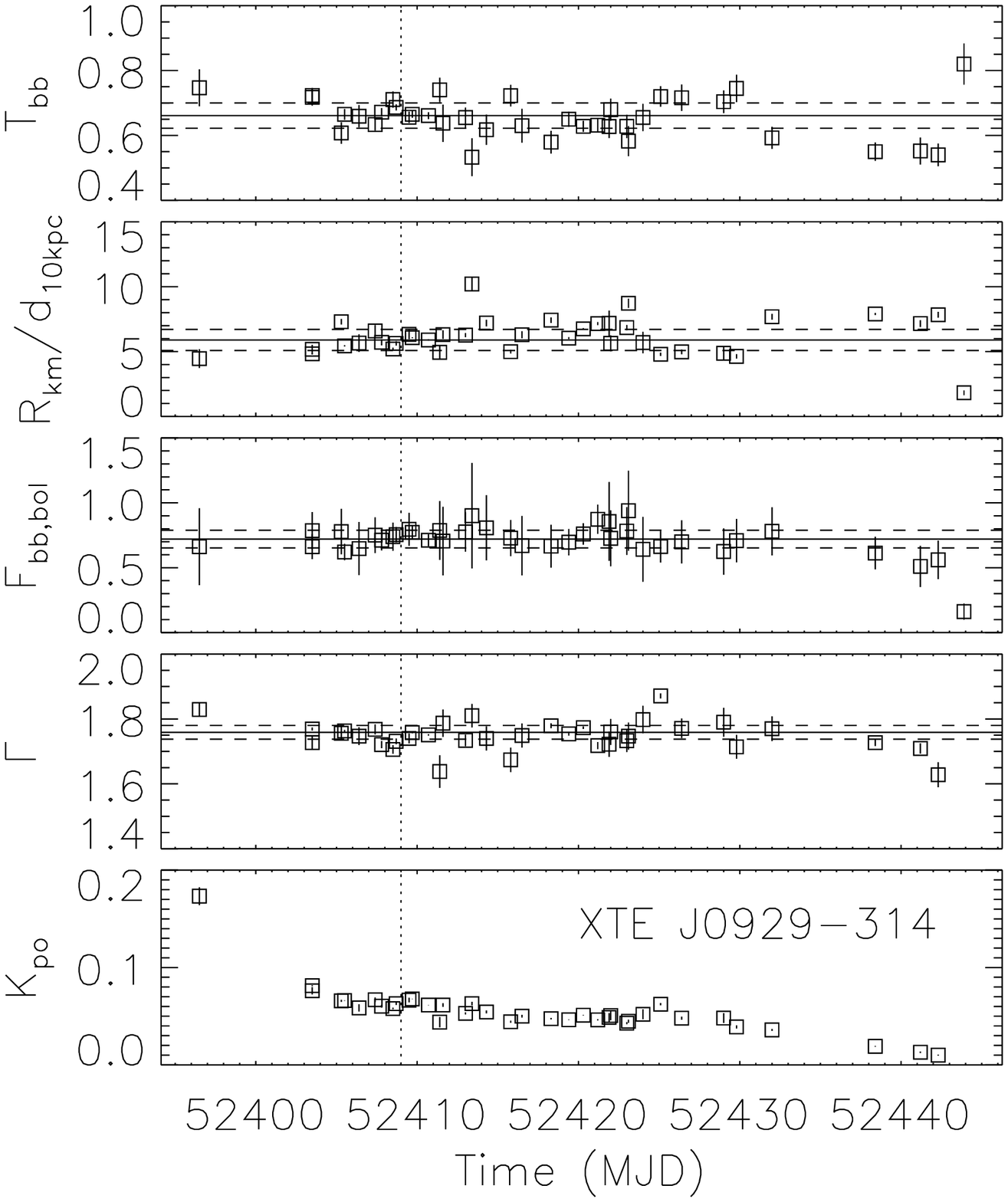,width=0.95\linewidth}} 
\figcaption{Spectral fit parameters derived from fitting an absorbed
blackbody $+$ power-law model to {\it RXTE}\/ observations of
XTE~J0929$-$314. The panels, from top to bottom, show the blackbody
temperature $T_{\rm bb}$ in keV; the blackbody normalization ($=R_{\rm
km}/d_{\rm 10kpc}$), estimated bolometric flux of the blackbody
component ($10^{-10}\ \epcs$), power-law photon index $\Gamma$, and
power-law normalization (photons$\,{\rm keV^{-1}\,cm^{-2}\,s^{-1}}$ at
1~keV). The solid line shows the weighted mean over all the
observations, while the dashed line shows the $\pm1\sigma$ limits. The
dotted line shows the time of the {\it Chandra}\/ observation.  Error
bars show the 1$\sigma$ uncertainties.}
\label{fig:1}
\vspace*{0.2in}

\section{Spectral Analysis}
\vspace{0.1in}

\subsection{\XTE\/ Spectral Analysis}
We fitted the combined 2--25~keV PCA and 15--50~keV HEXTE spectra with
a number of models which typically give good fits for other X-ray
pulsars, including broken and cutoff power laws and analytical
Comptonization approximations.  There was no evidence for a
high-energy spectral cutoff within the energy range in which the
source was detected (typically $\la50$~keV). We did however measure
significant residuals below 10~keV. These residuals could be minimized
by adopting a broken power law model where the spectral index
decreased by $\approx10$\% (i.e. the spectrum became harder) above
7~keV, or by adding a blackbody component with $kT_{\rm
bb}\approx0.5$~keV to a power law.  While both these models gave
fits of similar quality, we rejected the former as unphysical and
instead adopted the latter for the remainder of the \XTE\/ fits.

We also included a multiplicative component to take into account the
attenuation by intervening (neutral) material with cosmic abundances.
The equivalent hydrogen column density $N_{\rm H}$ exhibited no
significant variations between observations, and in the mean was
$\sim10^{21}\ {\rm cm}^{-2}$, consistent with the line-of-sight values
interpolated from dust and \ion{H}{1} maps \citep*[$7.6\times10^{20}$
and $10^{21}\ {\rm cm}^{-2}$ respectively;][]{sfd98,dl90}. For
consistency with the fits to \Ch\/ data, we froze the $N_{\rm H}$
value in subsequent fits at $7.6\times10^{20}\ {\rm cm}^{-2}$.  The
mean value of the reduced-$\chi^2$ ($=\chi^2_\nu$) for power-law $+$
blackbody fits to spectra from all the observations was 1.04.  While
the worst-fitting observation gave a $\chi^2_\nu=2.44$, the largest
residuals arose from variations between PCUs (particularly at the
lower energy bound) rather than any broad trend with energy. By
excluding these lowest channels from the fit, and assuming a
systematic error of 1\%, we were able to reduce the fit statistic for
that observation to $\chi^2_\nu=1.30$.

The weighted mean of the blackbody temperature, $kT_{\rm bb}$, was
$0.66\pm0.06$~keV, while the normalization $\sqrt{K_{\rm bb}}=R_{\rm
bb,km}/d_{\rm 10kpc}=5.9\pm1.3$ (all errors are quoted at
90\%-confidence unless otherwise noted).  Both parameters showed
variations during the outburst; however, they were significantly
anticorrelated, so that the estimated bolometric blackbody flux
($\propto K_{\rm bb}T_{\rm bb}^4$) remained constant to within our
measurement uncertainties ($\chi^2_\nu=1.06$, see Figure~1).  In the
first observation, the blackbody component comprises only 3\% of the
flux in the 2--60~keV band, whereas in the last observation where
significant blackbody and power-law components were detected, the
blackbody accounted for 16\% of the flux.

We calculated a weighted mean photon index of $\Gamma=1.76\pm0.03$.
The principal source of flux variation was the normalization of the
power-law component (see Figure~1).  While all parameters exhibited
variations throughout the outburst, the power-law normalization was
the only parameter to show a significant correlation with the
integrated 2--10~keV flux \citep[Spearman's rank correlation
coefficient $\rho=0.908$, equivalent to 5.4$\sigma$;][]{ptv+92}.  A
power-law component was not significantly detected in the last
observation on MJD~52443.  A blackbody-only fit for that day exhibits
a significantly (6.6$\sigma$) lower integrated blackbody flux than the
weighted mean level for the previous observations.  We did not detect
\J\/ in subsequent observations (3$\sigma$ upper limit on the
2--10~keV flux of $7.5\times10^{-12}\ \epcs$).

\subsection{\Ch\/ Spectral Analysis}
We fit the \Ch\/ spectrum with a power-law $+$ blackbody model
including absorption.  The $+$1\farcm5 Y-offset in the observational
setup allows for the entire 1.5--25.3 \AA\/ (0.5--8.3 keV) $+1$ order
spectrum to be imaged on the S3 CCD.  This CCD is backside-illuminated
and has suffered less degradation than the front-side illuminated
CCDs.  Because of this and the lack of counts above 25~\AA, we
performed our spectral fits on only the +1 order spectrum in the range
1.5--25.3 \AA.

Of the three known X-ray MSPs, \J\/ has the highest Galactic latitude
and, probably the smallest interstellar column density along the line
of sight to the source.  Our \Ch\/ observation allows for a direct
determination of the strength of the absorption edges, in particular
the O-K edge and the Fe-L triplet.  To account for absorption and
measure the optical depths of the edges, we used the {\tt tbvarabs}
absorption model \citep*[see,][]{wam00} with O and Fe abundances set
to zero and the $N_{\rm H}$ fixed to 7.6$\times10^{20}$~cm$^{-2}$, the
hydrogen column from dust maps \citep{sfd98}.  The O edge was fit with
an {\tt edge} model, while the Fe-L edges were fit using a custom
model that employs the optical constant measurements of \citet{kk00}.
In addition, the C abundance in {\tt tbvarabs} was allowed to vary in
order to account for the known instrumental contamination of the ACIS
CCDs\footnote{For more information see,
http://cxc.harvard.edu/cal/Links/Acis/acis/Cal\_prods/qeDeg/index.html}
\citep{psm+02}.  We also included Gaussian lines to fit the
interstellar atomic O absorption line at 23.5 \AA\/ and the absorption
feature  

\centerline{\epsfig{file=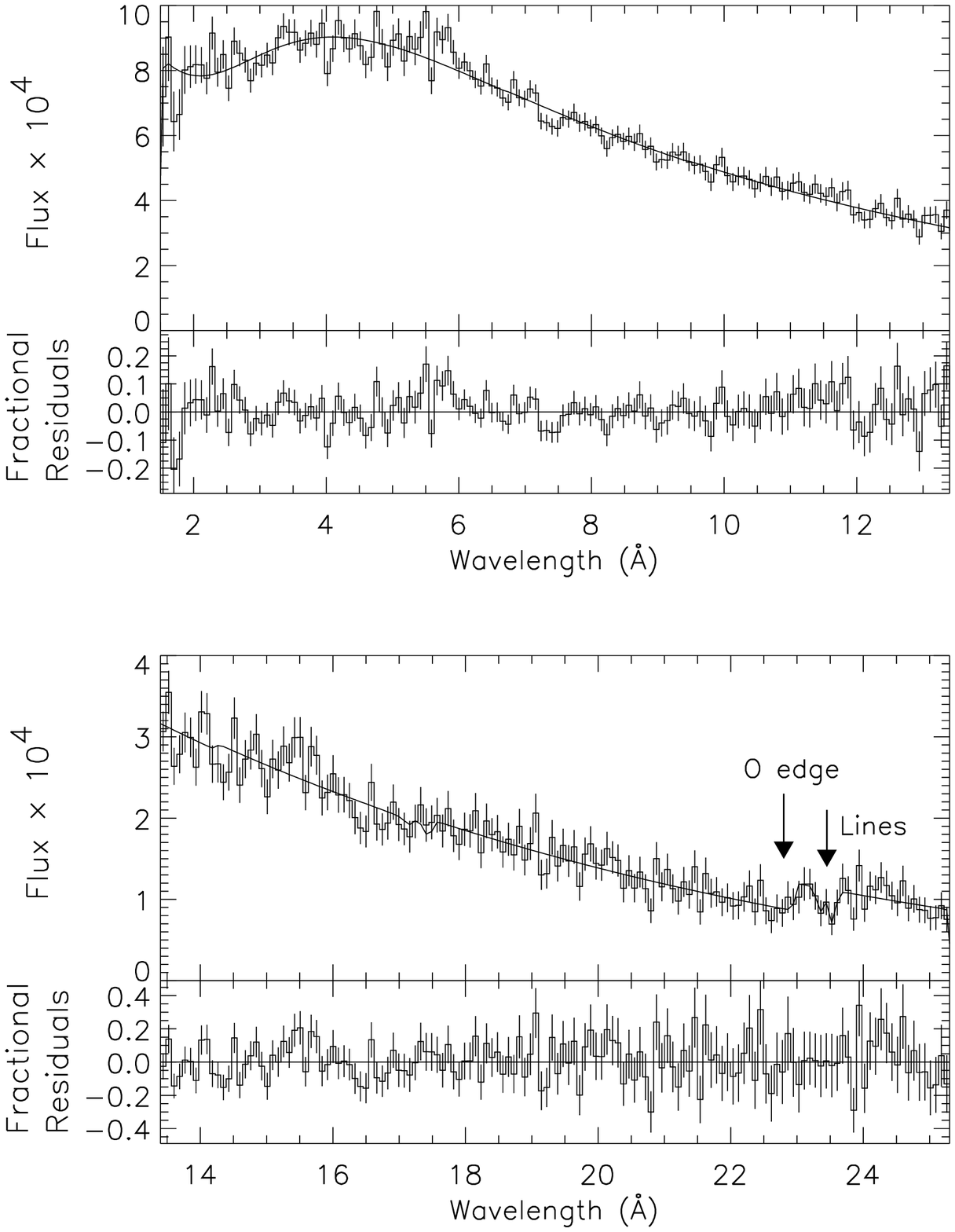,width=0.9\linewidth}} 
\figcaption{({\em upper panels}) \Ch\/ LETG $+1$ order flux spectrum
(in units of photons~cm$^{-2}$~s$^{-1}$~\AA$^{-1}$) of \J\/ with
best-fit power-law $+$ blackbody model with absorption.  The arrows
indicate the positions of the O edge, and the atomic O and
Fe$_{2}$O$_{3}$ absorption lines. ({\em lower panels}) Fractional
residuals ([data$-$model]/model) of the LETG spectral fit shown above.
The absorption models are consistent with the predicted interstellar
absorption from dust maps.}
\label{fig:2}
\vspace*{0.2in}

\noindent at 23.36 \AA, attributed to a 1s-2p transition in
Fe$_{2}$O$_{3}$ \citep[see,][]{scc+02}.

We found a best-fit photon index of 1.55$\pm$0.03, with a
normalization of (4.04$\pm$0.15)$\times 10^{-2}$ photons~keV$^{-1}$
cm$^{-2}$ s$^{-1}$ at 1 keV.  The \Ch\/ best-fit blackbody temperature
was $kT_{\rm bb}=0.65\pm0.03$~keV, with $R_{\rm bb,km}/d_{\rm
10kpc}=7.6\pm0.8$.  Our best-fit model had a $\chi^2_\nu=1.001$ and an
unabsorbed 2--10 keV flux of 2.7$\times10^{-10}$ erg cm$^{-2}$
s$^{-1}$.

The best-fit optical depth for O was 0.34$\pm$0.07, which translates
to an equivalent hydrogen column density of (1.2$\pm$0.3)$\times
10^{21}$ cm$^{-2}$, using the cross-section of \citet*{hgd93} and the
O ISM abundance of \citet{wam00}.  Although this is somewhat higher
than the measured hydrogen column from dust maps, an instrumental
contribution to the O edge has been reported, with an optical depth of
$\approx$0.10 \citep{psm+02}.  This contribution would add $N_{\rm
H}\approx0.36\times 10^{21}$ cm$^{-2}$ to the expected value, making
the \Ch\/ O edge measurement compatible with the dust map measurement.
The best-fit line wavelength and equivalent width (EW) of the atomic O
absorption line were 23.52$\pm$0.05~\AA\/ and
$0.053^{+0.005}_{-0.03}$~\AA.  The Fe$_{2}$O$_{3}$ absorption line and
Fe-L edges were not significantly detected, with an upper limit on the
Fe column density of $3\times 10^{16}$~cm$^{-2}$, equivalent to an
$N_{\rm H}<1.1\times 10^{21}$~cm$^{-2}$.  The best-fit C abundance was
$7.8_{-0.7}^{+1.2}$ times the interstellar C/H ratio of \citet{wam00}.
This is slightly lower than the expected instrumental $+$ interstellar
contribution of 9.9$\pm$0.5 times the interstellar ratio.  The
instrumental edge depths for C and O were estimated using the recent
calibration results presented at the {\em Chandra X-ray Center}
website\footnotemark[5] \citep[see also,][]{psm+02}.

Besides the interstellar O absorption line, there were no prominent
emission or absorption lines in the high-resolution spectrum of \J\/
(see Figure 2).  We performed a careful search of the \Ch\/ spectral
residuals to place limits on the presence of any spectral features.
Gaussian models with fixed FWHM $=2000$ km s$^{-1}$ were fit at each
point in the wavelength range 1.5--25.3 \AA.  From this, we can place
a 3$\sigma$ upper limit on the EW of any line feature, either emission
or absorption.  The EW limit increases approximately linearly with
wavelength, varying from 0.007~\AA\/ at 1.5~\AA\/ to 0.12~\AA\/ at
24~\AA.

Our measured C abundance depends on the assumed continuum model since
the edge depth can not be directly measured (the C edge at 43~\AA\/ is
outside our bandpass).  In order to test if the C abundance was the
cause of the difference in the photon index between the \Ch\/ and
\XTE\/ fits, we refit the \Ch\/ spectrum with a fixed C abundance of
9.9 times the interstellar C/H ratio.  With the C abundance fixed, the
best-fit photon index was 1.62$\pm$0.03.  This is still significantly
different from the best-fit photon index found in the \XTE\/ spectra.
The best-fit power-law normalization increased to
(4.50$\pm$0.07)$\times10^{-2}$ photons~keV$^{-1}$ cm$^{-2}$ s$^{-1}$.
The other fit parameters were consistent with the previous fit within
errors.  We also considered the possibility that the pileup model was
giving rise to the lower photon index in the \Ch\/ spectral fits.
When the \Ch\/ data is fit without the pileup model, and excluding the
piled-up region (1--3~keV) of the spectrum, we find spectral
parameters consistent with the results found using the pileup model.

\vspace{0.2in}
\section{Combined \Ch\/ and \XTE\/ Spectral Fits}
To investigate the difference in the spectral results between \Ch\/
and \XTE, we performed combined spectral fits with the \Ch\/ and
\XTE\/ data.  A single \XTE\/ observation (70096-03-05-00), taken
within a few hours of the \Ch\/ data, was used to represent the \XTE\/
spectrum of \J.  To simplify fitting, the piled-up region (1-3 keV) of
the \Ch\/ data was excluded and the pileup kernel was not used.  The
combined data were fit with a blackbody $+$ either a power-law, broken
power-law or the {\tt comptt} model which describes Comptonization of
soft photons by a hot plasma \citep{t94}.  The broken power-law and
Comptonization models allow for turnover at low energies which could
explain the difference in the \Ch\/ and \XTE\/ spectral results.  The
absorption was described by the {\tt tbvarabs} model with the C
abundance allowed to vary for the \Ch\/ observation only, while
best-fit edge depths from the \Ch\/ analysis were used for the O and
Fe edges.  We also included a Gaussian line, fixed to the best-fit
\Ch\/ value, to model the atomic O absorption at 23.5 \AA.
Multiplicative constants were included to compensate for any
instrumental normalization differences.  We find that the
normalizations of the PCA spectra are $\approx$20\% larger than for
\Ch, while the HEXTE spectra are $\approx$20\% lower.  Similar
normalization differences have been noted before \citep[see,
e.g.,][and references therein]{khz+02}.

\begin{figure*}[b]
\centerline{\epsfig{file=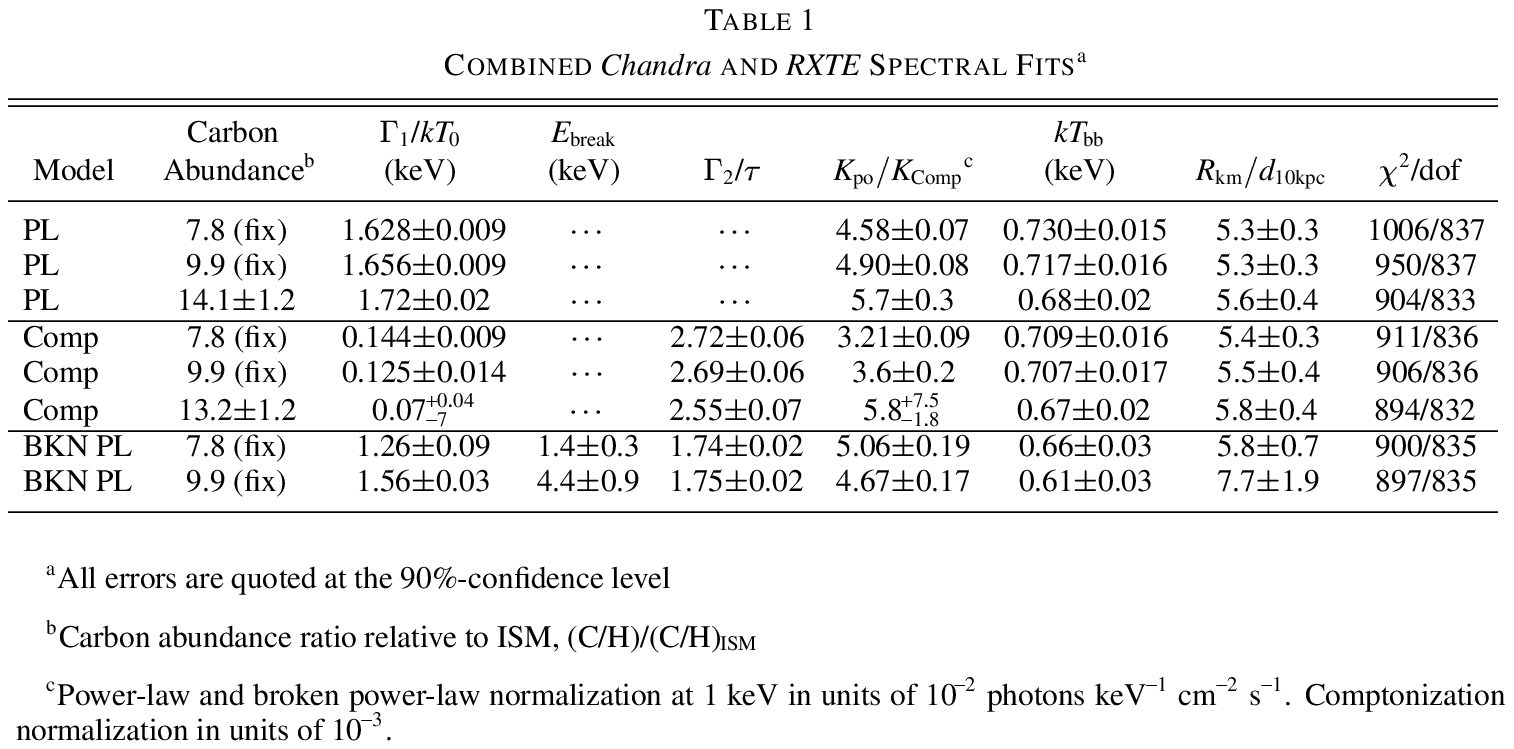}}
\end{figure*}

As a check on the instrumental calibrations, we directly compared the
\Ch\/ and \XTE\/ data in the same energy range by fitting the high
energy (3--8 keV) \Ch\/ data, in conjunction with the \XTE\/ data, to
an absorbed power-law $+$ blackbody model with absorption fixed to
$N_{\rm H}=7.6\times10^{20}$ cm$^{-2}$.  The fits are consistent with
the results of the \XTE\/ data, $\Gamma=1.74\pm0.02$, $kT_{\rm
bb}=0.66\pm0.03$~ keV, and $R_{\rm bb,km}/d_{\rm 10kpc}=5.9\pm0.7$,
with $\chi^2_\nu=1.03$.

We then fit the full \Ch\/ and \XTE\/ spectra with a power-law $+$
blackbody model.  When the edge parameters were fixed to the \Ch\/
best-fit values, the photon index had a best-fit value of
1.628$\pm$0.009 and with a $\chi^2_\nu=1.20$ (see Table 1).  We note
that the $\chi^2_\nu$ of the power-law $+$ blackbody fit to the \XTE\/
data alone was 1.06.  If we fix the C abundance to the expected value
of 9.9 times the interstellar C/H ratio, the photon index increases to
1.656$\pm$0.009 ($\chi^2_\nu=1.14$).  If we instead allow the C
abundance to vary, the best-fit photon index increases to
1.72$\pm$0.02 ($\chi^2_\nu=1.09$), giving a photon index consistent
with the \XTE\/ spectral results, but with a C abundance of
14.1$\pm1.2$ times the interstellar C/H ratio.  This value is
inconsistent with the estimate of the C abundance from instrument
calibration work (see \S 4.2).  As can be seen, the continuum spectral
parameters are highly dependent on the assumed absorption.

We performed the same fits using the {\tt comptt} in place of the
power-law component.  The Comptonization model provides for a
low-energy turnover in the spectrum.  The electron temperature was
fixed to 50 keV, the upper limit of the energy range.  In addition, we
used a spherical geometry in order to compare with the spectral fit of
SAX~J1808.4$-$3658 \citep*{tcw02}.  The {\tt comptt} model provides a
better fit to the data compared with the power-law
($\chi^2_\nu=1.09$--1.07, see Table 1).  We performed an $F$-test to
formally compare the two models and found that in all cases the
Comptonization model was better with a significance $>99$\%.  For the
two fixed values of the C abundance, the results of the {\tt comptt}
$+$ blackbody fits are consistent within error with a seed photon
temperature, $kT_0\approx0.13$~keV and optical depth,
$\tau_{p}\approx2.7$.  When the C abundance is allowed to vary, the
fit improves but becomes inconsistent with the other results, which is
probably due to degeneracy between the parameters.

In the best-fit {\tt comptt} $+$ blackbody model, the blackbody
component has parameter values of $kT_{\rm bb}=0.71$~keV and $R_{\rm
bb,km}/d_{\rm 10kpc}=5.5$.  These values are consistent with the
power-law fits both to the combined data and with the blackbody
parameters from each instrument fit, indicating that the blackbody
temperature and flux are independent of the exact continuum model
used.  We find $kT_{0}<kT_{\rm bb}$, similar to the fit of
SAX~J1808.4$-$3658 using the same model \citep{tcw02}.  We note that
the {\tt comptt} model is in itself an approximation to a full
treatment of Comptonization, in particular at soft energies.  If we
assume that the blackbody provides the input photons to the
Comptonization component, then we would expect $kT_{0}=kT_{\rm bb}$.
We fit the data with the constraint $kT_{0}=kT_{\rm bb}$, but the fit
was unreasonable with $\chi^2_\nu>2.0$.  Interestingly, \citet*{gdb02}
used a different Comptonization model and found $kT_{0}=kT_{\rm bb}$
in spectral fits of SAX~J1808.4$-$3658.  We employed the same model to
fit \J\/ \citep[{\tt compPS};][]{ps96}, but found similar values for
the input photon temperature and optical depth as found using the {\tt
comptt} model.

We also fit the data with a broken power-law $+$ blackbody model to
test for low-energy turnover without relying on the assumptions
inherent in the Comptonization model.  We find that the data is also
well fit ($\chi^2_\nu=1.08$) by a broken power-law $+$ blackbody model
with $\Gamma_1=1.26$--1.56, $\Gamma_2=1.75$, and $E_{\rm
break}=1.4$--4.4~keV, and blackbody parameters consistent with
previous fits (see Table 1).

\vspace{0.2in}
\section{Discussion}
We have found that the spectrum of \J\/ is well described by a
power-law $+$ blackbody model over limited energy ranges, with
interstellar absorption consistent with $N_{\rm H}=7.6\times
10^{20}$~cm$^{-2}$, the expected hydrogen column density along the
line of sight to the source.  The \Ch\/ spectrum shows no prominent
emission or absorption lines in the spectrum \citep[similar to
XTE~J1751$-$305;][]{mwm+02}, with an EW limit which varies from
$<0.007$~\AA\/ at 1.5~\AA\/ to $<0.12$~\AA\/ at 24~\AA.  No orbital
modulation of the X-ray flux was detected in either the \Ch\/ or
\XTE\/ data.  From the lack of X-ray eclipses, we set an upper limit
on the binary inclination of $i\lesssim 85^{\circ}$ for a
Roche-lobe--filling companion.  In addition, the absence of dips in
the X-ray lightcurve suggests that the source is most likely at an
inclination of $i\lesssim60^{\circ}-70^{\circ}$ \citep*[see,
e.g.,][]{fkl87,wnp95}.

The blackbody spectral component in LMXBs is generally attributed to
the neutron star, possibly from a ``hot spot'' where the accretion
column meets the neutron star surface.  From our fits, we find a
blackbody radius of $R_{\rm bb,km}=(5.3-7.7)\ d_{\rm 10kpc}$.  Given
the lower limit on the distance of 5~kpc \citep{gcm+02}, we find a
lower limit on the blackbody radius of $R_{\rm bb}>2.7$~km.  We note
that this lower limit does not include a correction for the conversion
between the color temperature and the effective temperature of the
neutron star atmosphere \citep*[see,][for a discussion]{lvt93}.  This
correction could increase the inferred radius by a factor of
$\approx2$.  But even after this correction, the inferred radius is
not consistent with canonical NS models for a distance of 5~kpc
suggesting a ``hot spot'' emission region.

During the outburst, we found that the blackbody flux remained
constant, while the power-law flux declined.  In addition, when the
combined \Ch\/ and \XTE\/ spectrum is fit with a Comptonization model,
we find that the blackbody temperature, $kT_{\rm bb}$, is
significantly greater than the seed photon temperature of the
Comptonization model, $kT_0$.  This evidence suggests that the
power-law and blackbody components are associated with different
emission processes.

We compared our results with two physical models that were developed
to explain the spectral properties of the accretion-powered MSP
SAX~J1808.4$-$3658 \citep{gdb02,tcw02}.  Both models were formulated
using data from the 1998 outburst of SAX~J1808.4$-$3658, but come to
very different conclusions regarding the relationship between the
power-law or Comptonization component and the blackbody.
\citet{gdb02} assumed that the blackbody was the source of the seed
photons (i.e, $kT_0 = kT_{\rm bb}$) and proposed that the
Comptonization component comes from the shock heated accretion column
with soft photon input from the ``hot spot'' blackbody component.
Such a model implies that the blackbody and power-law/Comptonization
component fluxes should be correlated, which was the case for
SAX~J1808.4$-$3658 but was not seen for \J.

On the other hand, \citet{tcw02} found that $kT_0 << kT_{\rm bb}$ and
suggested that the accretion disk provided the majority of the soft
photon input to a spherical Compton cloud, of radius $R\approx~60$~km,
from which the Comptonization component originates.  The \citet{tcw02}
model allows for the the blackbody and power-law/Comptonization
component to vary independently. \citet{tcw02} also suggested that the
low ($\tau_{0}\approx4$) optical depth of the Comptonizing region
found for SAX~J1808.4$-$3658 allows for the pulsations of the NS to be
detected, while the higher optical depths ($\tau_{0}>5$) reported in
more luminous LMXBs suppress the pulsation amplitudes.  We find a
best-fit $\tau_{0}\approx2.7$ for \J, consistent with this
explanation.  However, we note that \citet{mwm+02} did not find an
acceptable fit of the {\em XMM} spectra of the millisecond X-ray
pulsar XTE~J1751$-$305 using a Comptonization model.  While this model
seems more appropriate for the spectral results of \J, \citet{tcw02}
do not discuss what emission component is responsible for the
pulsations.

It is important to realize that both models are based on data with a
low-energy bound of 2~keV, which is above the peak flux of the thermal
components.  The lower energy range of \Ch\/ allows us to make a more
robust measurement of $kT_0$ for \J\/ than was possible for the \XTE\/
spectrum of SAX~J1808.4$-$3658.  We conclude that neither model is
completely appropriate to explain both our results.  More observations
of these sources throughout outburst are necessary in order to fully
understand their phase-averaged spectral properties.  We suggest that
in-depth data analysis, including pulse-phase spectroscopy, will
provide more insight into the relationship between the various
spectral components.  The pulse-phase resolved spectral analysis of
XTE~J0929$-$314 will be presented in a later paper.

We find that the best-fit photon index was significantly different
between the \Ch\/ and \XTE\/ fits with $\Gamma=1.55-1.62$ for the
\Ch\/ fits, dependent on the assumed instrumental C abundance, and
$\Gamma=1.76$ for the \XTE\/ fits.  Since fits with and without the
pileup model give consistent parameter values, we reject the
possibility that pileup causes the lower \Ch\/ photon index.  The
combined \Ch\/ and \XTE\/ spectrum of \J\/ is well fit by either a
Comptonization $+$ blackbody model (which implicitly includes a
low-energy turnover from the Comptonization component), or a broken
power-law $+$ blackbody model, with break energy 1.4--4.4~keV.  The
high-energy ($E>3$~keV) \Ch\/ spectrum is consistent with the \XTE\/
spectral results, indicating that instrumental differences alone do
not give rise to the spectral turnover.  Based on these results, we
suggest that the difference in best-fit photon indices between \Ch\/
and \XTE\/ arises from spectral turnover at low energies.  While a
power-law model is a reasonable approximation to Comptonization at
high energies, extrapolating the power law to low energies leads to a
predicted flux that diverges at zero energy.  Therefore, the power-law
must turnover at some energy for the integrated to flux to be finite.

In addition, the difference in the photon index between the \Ch\/ and
\XTE\/ spectral fits is similar to that noted by \citet{mwm+02}
between the \XTE\/ and {\em XMM} spectral fits of XTE J1751$-$305.
From their {\em XMM} EPIC spectrum, \citet{mwm+02} found a best-fit
power-law photon index of 1.44, while \citet{mss+02} found a best-fit
photon index of 1.7--1.9 using \XTE.  We interpret the difference
between the {\em XMM} and \XTE\/ spectral results for XTE J1751$-$305
as evidence of low energy turnover in the broadband spectrum of that
source, providing independent support that the turnover is
astrophysical in origin.

\acknowledgments{We would like to thank \Ch\/ Director Harvey
Tananbaum and \XTE\/ Director Jean Swank for approving our DDT/TOO
observations.  In addition, we acknowledge useful discussions with
Craig Markwardt, Jon Miller, Mike Nowak, and Dimitrios Psaltis.  This
research has made use of data obtained through the High Energy
Astrophysics Science Archive Research Center Online Service, provided
by the NASA/Goddard Space Flight Center.  This work was supported in
part by NASA under contract NAS8-01129 and grant NAG5-9184.}

\end{document}